\documentclass[11pt, letterpaper]{article} 
\usepackage[ margin=1.0in]{geometry}
\usepackage{microtype}
\usepackage{graphicx}
\usepackage{layout}
\usepackage{lipsum}
\usepackage[utf8]{inputenc}
\usepackage{fancyhdr}
\usepackage{xspace}
\usepackage{tabularx}
\usepackage{algorithm}
\usepackage{algorithmicx}
\usepackage{algcompatible}
\usepackage{algpseudocode}
\algrenewcommand\algorithmicindent{0.3em}%
\usepackage[toc,page]{appendix}
\usepackage{color}
\usepackage{caption}
\usepackage{ragged2e}
\usepackage{tabularx}
\usepackage{amsmath,amssymb,latexsym} 
\usepackage{fancyvrb}
\usepackage{float}
\usepackage{cite}
\usepackage{epstopdf}
\usepackage{epsfig}
\usepackage{graphicx}
\usepackage{pgfplots}
\usepackage{lipsum}
\usepackage{booktabs,chemformula}
\usepackage{tabularx,booktabs} 
\usepackage{etoolbox}
\usepackage{breqn}
\usepackage{subfiles}
\usepackage{sidecap}
\usepackage{authblk}
\usepackage{soul}
\usepackage{mathtools}
\usepackage{pdfpages}
\pgfplotsset{compat=1.7}
\usepackage{calligra}
\usepackage{calrsfs}
\DeclareMathAlphabet{\mathcalligra}{T1}{calligra}{m}{n}
\DeclareCaptionFont{tiny}{\tiny}
\algrenewcommand\alglinenumber[1]{\tiny #1:}
\usepackage{multicol}
\usepackage{comment}
\usepackage{subcaption}
\usepackage{authblk}
\usepackage{appendix}
\usepackage{xspace}
\usepackage{enumitem}
\usepackage{soul}
\usepackage{mathtools}
\usepackage{breqn}
\usepackage{hyperref}
\usepackage{url}
\hypersetup{
    colorlinks=true,
    linkcolor=blue,
    filecolor=magenta,      
    urlcolor=cyan,
}
\newcommand{\punt}[1]{}
\newcommand{\cmnt}[1]{}

\newtheorem{theorem}{Theorem}

\newtheorem{lemma}[theorem]{Lemma}

\newtheorem{definition}{Definition}

\newcounter{history}

\newcommand{\secref}[1]{Section~\ref{sec:#1}}

\newcommand{\stref}[1]{Step~\ref{step:#1}}

\newcommand{\lineref}[1]{Line~\ref{lin:#1}}

\newcommand{\ignore}[1]{}

\newcommand{\mth} {method\xspace}

\newcommand{\sspec} {sequential\text{-}specification\xspace}

\newcommand{\lble} {linearizable\xspace}
\newcommand{\lbty} {linearizability\xspace}

\newcommand{\cgds} {concurrent graph data structure\xspace}

\newcommand{\lp} {LP\xspace}

\newcommand{\sgt} {SGT\xspace}

\newcommand{\vnodes}[1] {#1.vnodes\xspace}
\newcommand{\enodes}[1] {#1.enodes\xspace}
\newcommand{\enode}{enode\xspace}
\newcommand{\vnode}{vnode\xspace}
\newcommand{\vlist} {vertex\text{ }list\xspace}

\newcommand{\helpe}{HelpSearchEdge\xspace}

\newcommand{\vh}{VertexHead}
\newcommand{\vt}{VertexTail}
\newcommand{\eh}{EdgeHead}
\newcommand{\et}{EdgeTail}
\newcommand{\abg} {AbG\xspace}
\newcommand{\abgv} {AbG(V)\xspace}
\newcommand{\abge} {AbG(E)\xspace}
\newcommand{\addv}{AddVertex\xspace}
\newcommand{\remv}{RemoveVertex\xspace}
\newcommand{\adde}{AddEdge\xspace}
\newcommand{\reme}{RemoveEdge\xspace}
\newcommand{\conv}{ContainsVertex\xspace}
\newcommand{\cone}{ContainsEdge\xspace}

\newcommand{\add}{Add\xspace}
\newcommand{\rem}{Remove\xspace}
\newcommand{\con}{Contains\xspace}

\newcommand{\acadde}{AcyclicAddEdge\xspace}

\algrenewcommand{\algorithmiccomment}[1]{$//$ #1}

\newcommand{\concgraph}{ConcGraph\xspace}
\newcommand{\ds}{data-structure\xspace}
\newcommand{\fg}{fine-grained\xspace}
\newcommand{\cg}{coarse-grained\xspace}
\newcommand{\lf}{lock-free\xspace}
\newcommand{\wf}{wait-free\xspace}

\newcommand{\naddv}{AddVertex\xspace}

\newcommand{\nreme}{RemoveEdge\xspace}
\newcommand{\wconv}{ContainsVertex\xspace}
\newcommand{\wcone}{ContainsEdge\xspace}
\algrenewcommand{\algorithmiccomment}[1]{$//$ #1}

\newcommand{\node}{node}

\newcommand{\head}{Head\xspace}

\newcommand{\loct}{Locate\xspace}
\newcommand{\valid}{Validate\xspace}
\newcommand{\lfadd}{LFAdd\xspace}
\newcommand{\lfrem}{LFRemove\xspace}

\newcommand{\lfloct}{LFLocate\xspace}

\newcommand{\wfcon}{WFContains\xspace}

\newcommand{\lfaddv}{LFAddVertex\xspace}
\newcommand{\lfremv}{LFRemoveVertex\xspace}
\newcommand{\lfadde}{LFAddEdge\xspace}
\newcommand{\lfreme}{LFRemoveEdge\xspace}
\newcommand{\wfconv}{WFContainsVertex\xspace}
\newcommand{\wfcone}{WFContainsEdge\xspace}

\newcommand{\hnode}{node\xspace}

\newcommand{\hhead}{Head}

\newcommand{\hadd}{HoHAdd\xspace}
\newcommand{\hrem}{HoHRemove\xspace}
\newcommand{\hcon}{HoHContains\xspace}
\newcommand{\hloct}{HoHLocate\xspace}

\newcommand{\hoh}{hoh-locking-list\xspace}
\begin{document}
	
\title{\textbf{Building Efficient Concurrent Graph Object through Composition of List-based Set
}\footnote{ This work is currently in progress. }
}

\author{
      Sathya Peri, Muktikanta Sa$^*$, Nandini Singhal$^*$ \\
      Department of Computer Science \& Engineering \\
      Indian Institute of Technology Hyderabad, India \\
      \{sathya\_p, cs15resch11012, cs15mtech01004\}@iith.ac.in
}

\date{}

\maketitle

\begin{abstract}
 In this paper, we propose a generic concurrent directed graph (for shared memory architecture) that is concurrently being updated by threads adding/deleting vertices and edges. The graph is constructed by the composition of the well known concurrent list-based set \ds from the literature. Our construction is generic, in the sense that it can be used to obtain various progress guarantees, depending on the granularity of the underlying concurrent set implementation - either blocking or non-blocking. We prove that the proposed construction is linearizable by identifying its linearization points. Finally, we compare the performance of all the variants of the concurrent graph \ds along with its sequential implementation. We observe that our concurrent graph \ds mimics the performance of the concurrent list based set. 

\end{abstract}

\textbf{Keywords: }concurrent data structure; lazy set; directed graph; non-blocking; locks; lock-freedom;

\section{Introduction}
\label{sec:intro}
A graph represents pairwise relationships between objects along with their properties. Due to their usefulness, graphs are being used in various kinds of networks such as social, semantic, genomics, etc. Generally, these graphs are very \textit{large} and \textit{dynamic} in nature. Dynamic graphs are the one's which are subject to a sequence of changes like insertion, deletion of vertices and/or edges \cite{Demetrescu2010}. Online social networks (facebook, linkedin, google+, twitter, quora, etc.), are dynamic in nature with the users and the relationships among them changing over time. 
In this paper, we develop a generic concurrent directed graph data-structure, which allows threads to concurrently add/delete or perform contains on vertices/edges while ensuring \lbty\cite{HerlWing:1990:TPLS}. 
The graph is constructed by the composition of the well known concurrent list-based set implementation from the literature. To the the best of our knowledge, this is the first work to propose an adjacency list representation of a graphs by composing list-based sets. The other known work on concurrent graphs by Kallimanis \& Kanellou \cite{Kallimanis} works on adjacency matrix representation and assume an upper-bound on the number of vertices while we make no such assumption. Our construction is generic, in the sense that it can be used to obtain different progress guarantees, depending on the granularity of the underlying concurrent set implementation - either blocking or non-blocking. The blocking list implementation is taken from Heller et. al \cite{Heller-PPL2007} which is also popularly known as the lazy implementation whereas the non-blocking variant is by Harris \cite{Harrisdisc01}. Our design is not a straight forward extension of the concurrent list-based set implementation but has several non-trivial additions. This can be seen from the Linearization Points of edge update methods (described later) which lie outside their method code \& depend on other concurrently executing graph methods. We believe the design of the concurrent graph data-structure is such that it can help identify other useful properties on graph such as cycle detection, shortest path, reachability, minimum spanning tree, strongly connected components, etc.

\textbf{Roadmap} - In \secref{System-Model-Preliminaries}, we describe the system model \& preliminaries detailing definitions of the terminology used in the paper. In \secref{con-graph-ds}, we describe the construction of our generic concurrent graph data structure. Later on, in \secref{working-con-graph-methods}, we describe the working of each of the methods and their linearization points. In \secref{simulation-results}, we describe the evaluation of the throughput of the several variants - sequential, using coarse-grained locking, hand-over-hand locking, lazy list-based set, lock-free implementation. Finally we conclude in \secref{conclusion-future-directions}. We also provide the implementation of the lazy list-based set and lock-free implementation in the Appendix \ref{sec:app-pcode}, for the readers' quick reference.

\vspace{-2mm}
\section{System Model \& Preliminaries}
\label{sec:System-Model-Preliminaries}
\vspace{-2mm}
In this paper, we assume that our system consists of finite set of $p$ processors, accessed by a finite set of $n$ threads that run in a completely asynchronous manner and communicate using shared objects. The threads communicate with each other by invoking \mth{s} on the shared objects and getting corresponding responses. Consequently, we make no assumption about the relative speeds of the threads. We also assume that none of these processors and threads fail. Our algorithm is designed for execution on a shared-memory multi-processor system which supports atomic $read$, $write$ and \emph{compare-and-swap(CAS)} operations. \\
\noindent
\textbf{Safety:} To prove a concurrent data-structure to be correct, \textit{\lbty} proposed by Herlihy \& Wing \cite{HerlWing:1990:TPLS} is the standard correctness criterion in the concurrent world. They consider a history generated by a data-structure which is collection of \mth invocation and response events. Each invocation of a method call has a subsequent response. A history is \lble if it is possible to assign an atomic event as a \emph{linearization point} (\emph{\lp}) inside the execution interval of each \mth such that the result of each of these \mth{s} is the same as it would be in a sequential history in which the \mth{s} are ordered by their \lp{s} \cite{MauriceNir}. \\
\ignore{
\textbf{Linearizability:} To prove a concurrent data-structure to be correct, \textit{\lbty} proposed by Herlihy \& Wing \cite{HerlWing:1990:TPLS} is the standard correctness criterion in the concurrent world. They consider a history generated by a data-structure which is collection of \mth invocation and response events. Each invocation of a method call has a subsequent response. A history to be \lble if (1) The invocation and response events can be reordered to get a valid sequential history. (2) The generated sequential history satisfies the object's sequential specification. (3) If a response event precedes an invocation event in the original history, then this should be preserved in the sequential reordering.  \\
}
\noindent
\textbf{Progress:} The \emph{progress} properties specifies when a thread invoking \mth{s} on shared objects completes in presence of other concurrent threads. Some progress conditions used in this paper are mentioned here which are based on the definitions in Herlihy \& Shavit \cite{opodis_Herlihy}. The progress condition of a method in concurrent object is defined as: (1) Blocking: In this, an unexpected delay by any thread (say, one holding a lock) can prevent other threads from making progress. (2) Deadlock-Free: This is a \textbf{blocking} condition which ensures that \textbf{some} thread (among other threads in the system) waiting to get a response to a \mth invocation will eventually receive it. (3) Lock-Free: This is a \textbf{non-blocking} condition which ensures that \textbf{some} thread waiting to get a response to a \mth (among multiple other threads), eventually receives it.
\vspace{-2mm}

\section{Construction of Concurrent Generic Graph Data-Structure}
\label{sec:con-graph-ds}
\begin{figure}[H]
		\centerline{\scalebox{0.43}{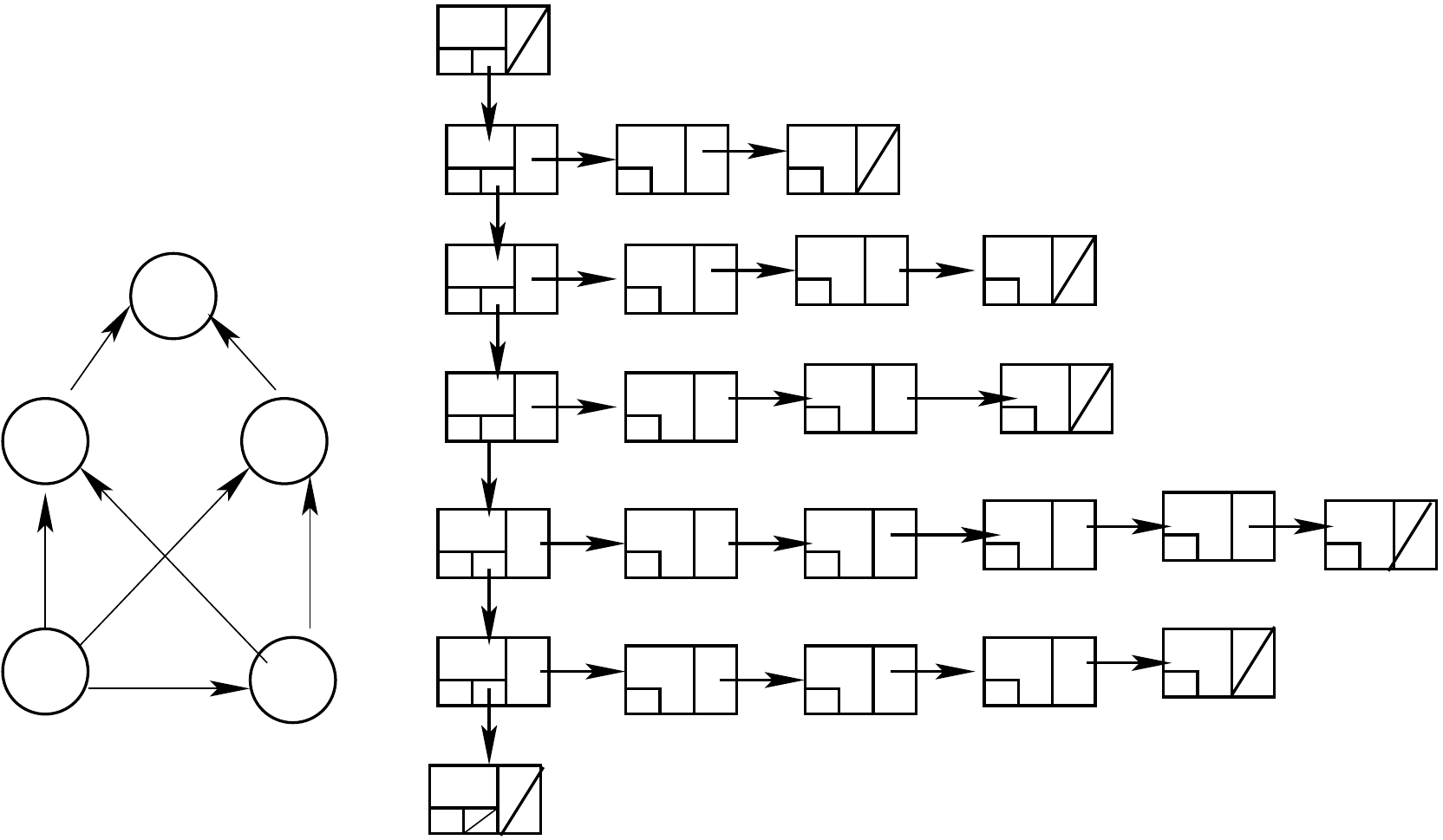}}
		\caption{(a) A directed Graph (b) The \cgds representation for (a).}
		\label{fig:conGraph}
	\end{figure}

In this section, we describe the construction of our generic concurrent graph \ds. We represent our \ds using the adjacency list representation of the graph. It is implemented as a list of lists, i.e., composition of lists. Each vertex in the list holds a list of vertices to which it has outgoing edges. This is depicted pictorially in Figure \ref{fig:conGraph}. \\

\noindent The problem addressed in this paper is described as follows: A concurrent directed graph $G = (V,E)$, where $G(V)$ is a set of vertices and $G(E)$ is a collection of directed edges. Each edge connects an ordered pair of vertices belonging to $G(V)$. And this $G$ is dynamically modified by a fixed set of concurrent running threads. In this setting, threads may perform insertion / deletion of vertices or edges to the graph. We assume that all the vertices have unique identification key (captured by $val$ field) as shown below.\\

\noindent In Table \ref{class:GNode}a, we describe the structure of the Node class. In the concurrent graph \ds, a \emph{GNode} could be used to represent either the vertex node or the edge node. Thus we add an additional next field to indicate the list we wish to traverse. In other words, if the \emph{GNode} is a vertex node, we will use its \emph{vnext} to point to the next vertex node and \emph{enext} to direct to its edge head sentinel. Similarly, if the GNode is a edge node, we will only use its enext to go to the next edge node in the edge list. The remaining fields of the GNode class are the same as the fields of concurrent linked list based on the lazy or the lock-free implementation. The construction shown here is keeping in mind the lazy \& lock free list-based set implementation which makes use of the additional marked field. However, it can be modified appropriately to obtain other variants.

\begin{multicols}{2}
	\scriptsize
	\vspace{-0.5cm}
	\label{class:GNode}
	\textbf{Table 3a}: Structure of GNode.
	\vspace{-0.2cm}
	\begin{Verbatim}[fontsize=\small]
	class GNode{
	int val;
	bool marked;
	GNode vnext;
	GNode enext;
	GNode(int key){  
	val = key;
	marked = false;
	vnext = enext = null;
	}
	};
	\end{Verbatim}
\end{multicols}
\noindent The Table \ref{class:List}b depicts the structure of the list-based set class. It creates two sentinel nodes for each list, called head and tail. A list can be uniquely identified using its head node. An object of the List class exports the add(x), remove(x) \& contains(x) methods. Each of these methods can be implemented using various well known synchronization techniques of the concurrent set implementation - fine-grained, optimistic, lazy or lock-free. The only difference is that we also pass the head of the list on which each of these operation should be performed. With our modified Node class, each GNode could possibly direct to either a vertex node or a edge node; we must identify correctly on which list we wish to perform the operation. To do this, we pass the head of the list we wish to operate on. If the argument passed is the head of the vertex list, we use the vnext to direct to the next GNode, otherwise enext.\\

\noindent Finally for our purposes, as illustrated in table \ref{class:Graph}c, the concurrent graph \ds exports the specified methods. These are explained in the next subsection.


\begin{multicols}{2}
	\scriptsize
	\label{class:List}
	\textbf{Table 3b}: Structure of List.
	\begin{Verbatim}[fontsize=\small]
	class List{
	GNode Head, Tail; 
	List(){
	//Initialize sentinels
	}
	bool Add(GNode Head, int key);
	bool Remove(GNode Head, int key);
	bool Contains(GNode Head, int key);
	/* performing operation on the list 
	represented by Head */
	};
	\end{Verbatim}
	\label{class:Graph}
	\textbf{Table 3c}: Structure of Graph.
	\begin{Verbatim}[fontsize=\small]
	class Graph extends List{
	bool AddVertex(int key);
	bool RemoveVertex(int key);
	bool ContainsVertex(int key);
	bool AddEdge(int key1, int key2);
	bool RemoveEdge(int key1, int key2);
	bool ContainsEdge(int key1, int key2);
	};
	
	Graph G;
	\end{Verbatim}
\end{multicols}



\ignore{
	\label{struct:vnode:enode}
	\textbf{Table \ref*{struct:vnode:enode}}: Structure of enode and vnode.
	\begin{verbatim}
	
	***Edge Node***    ***Vertex Node***
	class enode       class vnode
	{                 {
	int val;           int val; 
	enode enext;       vnode vnext;
	boolean marked;    enode EdgeHead;
	Lock lock;         enode EdgeTail; 
	};                   boolean marked; 
	Lock lock;
	};
	\end{verbatim}    
	\vspace{-0.6cm}
	\begin{figure}[H]
		\centerline{\scalebox{0.35}{\input{figs/conGraph.pdf_t}}}
		\caption{(a) A directed Graph (b) The \cgds representation for (a).}
		\label{fig:conGraph2}
	\end{figure}
}
\ignore{
	\begin{table*}[ht]
		\centering
		\scriptsize
		\noindent\begin{tabular}{ m{6cm}| m{6cm} } 
			\hline \\[-2ex]
			{Edge Node} &   {Vertex Node} \\ 
			\hrule & \hrule\\[-2ex]
			\begin{verbatim}
			class enode{
			/* unique value */
			int val; 
			/* pointer to the next enode */
			enode enext; 
			/* Marked indicates whether 
			this node is deleted or not. */
			boolean marked; 
			Lock lock; 
			};
			\end{verbatim}
			&
			\begin{verbatim}
			class vnode{
			/* a unique value */
			int val; 
			/* pointer to the next vnode */
			vnode vnext; 
			/* Head & Tail values initialized 
			to infinity */
			enode EdgeHead, EdgeTail; 
			/* Indicates whether this node is
			deleted or not */
			boolean marked; 
			Lock lock;
			};
			\end{verbatim}
			\hfill  \\[-2ex]
			\hrule & \hrule\\[-2ex]
		\end{tabular}
		\captionsetup{font=scriptsize}
		\caption{Structure of Vertex and Edge Node.}
		\label{tab:struct:vertex-edge}
		\vspace{-7mm}
	\end{table*}
}
\ignore{
	\noindent Given a global state $S$, we define a few structures and notations as follows: 
	\begin{enumerate}
		\item We denote vertex node, say $v$, as a $\vnode$ class object. Similarly, we denote edge node, say $e$, as a $\enode$ class object.
		\item $\vnodes{S}$ as a set of $\vnode$ class objects that have been created so far in $S$. Its structure is defined in the Table \ref{tab:struct:vertex-edge}. Each $\vnode$ object $v$ in $\vnodes{S}$ is initialized with $key$, $vnext$ to $null$, $marked$ field initially set to $false$. Two sentinel edge nodes are created: $\eh$ and $\et$ assigned with val $-\infty$ and $+\infty$ respectively; $\et.enext$ to $null$ and $\eh.enext$ to $\et$.
		\item $\enodes{S}$ is a set of \enode class objects and its structure is defined in the Table \ref{tab:struct:vertex-edge}.  Each $\vnode$ object $v$ in $\enodes{S}$ is initialized with $key$, $enext$ to $null$, with $marked$ field initially set to $false$.
		\item $S.\vh$ is a $\vnode$ class object (called sentinel head vertex node), which is initialized with a val $-\infty$. This sentinel node is never removed from the vertex list. Similarly, $S.\vt$ is a $\vnode$ class object (called sentinel tail vertex node), which is initialized with a val $+\infty$. This sentinel node is never removed from the vertex list.
		\item The contents \& the status of a $\vnode$ $v$ keeps changing with different global states. For a global state $S$, we denote $S.v$ as its current state and the individual fields of $v$ in $S$ as $S.v.val, S.v.vnext, ...$ etc.  
		\item Similar to the contents of a \vnode, it can be seen that the contents \& the status of a \enode $e$ keeps changing with different global states. Similarly, for a global state $S$, we denote $S.e$ as its current state and the individual fields of $e$ in $S$ as $S.e.val, S.e.vnext, ...$ etc.  
	\end{enumerate}
	
	\noindent Having defined a few notions on $S$, we now define the notion of an abstract graph, $\abg$ for a global state $S$ which we will use for guiding us in correctness of our \mth{s}. $\abg$ for a global state $S$ is the collection of $S.\abgv$ and $S.\abge$ as defined below:
	
	\begin{definition}
		\label{def:abgV}
		$S.\abgv \equiv \{v | (v \in \vnodes{S}) \land (S.\vh \rightarrow^* S.v) \land (\neg S.v.marked)\}$.
	\end{definition}
	
	\noindent This definition of \abgv captures the set of all vertices of $\abg$ for the global state $S$. It consists of all the \vnode{s} that are reachable from $S.\vh$ and are not marked for deletion. 
	
	\begin{definition}
		\label{def:abgE}
		$S.\abge \equiv \{e | (e \in S.enodes) \land ((u, v) \subseteq S.\abgv) \land (S.u.\eh \rightarrow^* S.e) \land (\neg S.e.marked) \land (S.e.val = S.v.val)\}$.
	\end{definition}
	
	\noindent This definition of $\abge$ captures the set of all edges of $\abg$ for the global state $S$. Informally it consists of all the $\enode$s that connects two $\vnode$s $u$, $v$ with the edge going form $u$ to $v$. 
	
}


\ignore{
	\noindent The problems addressed in this paper are defined as here: 
	\begin{enumerate}
		\vspace{-1mm}
		\item A concurrent directed graph $G = (V,E)$, which is dynamically modified by a fixed set of concurrent threads. In this setting, threads may perform insertion / deletion of vertices or edges to the graph. We develop this data-structure in Section \ref{sec:working-con-graph-methods}.
		\item We also maintain an invariant that the concurrent graph $G$ updated by concurrent threads should be acyclic. This means that the graph should preserve acyclicity at the end of every operation in the generated equivalent sequential history. We describe the modified data-structure in Section \ref{sec:Maintaining-Graph-Acyclicity}.  
	\end{enumerate}
}
\noindent
\vspace{-1cm}
\subsection{Methods Exported \& Sequential Specification}
\vspace{-1mm}
\noindent In this sub-section, we describe the \mth{s} exported by the concurrent directed graph data-structure  along with their sequential specification. The specification as the name suggests shows the behavior of the graph when all the \mth{s} are invoked sequentially. 
\vspace{-1mm}
\begin{enumerate}
	\item The $\addv(u)$ method adds a vertex $u$ to the graph, returning $true$. This follows directly from our assumption that all the vertices are assigned distinct keys. Once added, the method will never invoke addition on this key again.
	\vspace{-1mm}
	\item The $\remv(u)$ method deletes vertex $u$ from the graph, if it is present in the graph and returns true. By deleting this vertex $u$, this method ensures that all the incoming and outgoing edges of $u$ are deleted as well. If the vertex is not present in the graph, it returns $false$. 
	\vspace{-1mm}
	\item The $\conv(u)$ returns $true$, if the graph contains the vertex $u$; otherwise returns $false$.
	\vspace{-1mm}
	\item The $\adde(u, v)$ method adds a directed edge $(u, v)$ to the graph if the edge $(u, v)$ is not already present in the graph and returns $true$. If the edge is already in the graph it simply returns $true$. But if either the vertices $u$ or $v$ are not present, it returns $false$. 
	\vspace{-1mm}
	\item The $\reme(u,v)$ method deletes the directed edge $(u, v)$ from the graph structure if it is present and returns $true$. If the edge $(u, v)$ is not present in the graph but the vertices $u$ \& $v$ are in the graph it still returns $true$. But, if either of the vertices $u$ or $v$ are not present in the graph, it returns $false$. 
	\vspace{-1mm}
	\item The $\cone(u,v)$ returns $true$, if the graph contains the edge ($u,v$); otherwise returns $false$. 
\end{enumerate}
\vspace{-3mm}

\subsection{Sequential Specification}
\label{sec-app-seq-spec}
\noindent Given a global state $S$, we now define the notion of an abstract graph(for \fg, \lf and \wf, as \cg \& sequential are straight forward) , $\abg$ for a global state $S$ which we will use for guiding us in correctness of our \mth{s}. $\abg$ for a global state $S$ is the collection of $S.\abgv$ and $S.\abge$ as defined below:

\begin{table*}[!ht]
	\captionsetup{font=scriptsize}
	\centering
	\captionof{table}{Sequential Specification of the Graph Data-Structure } \label{tab:seq-spe} 
	\scriptsize
	\begin{tabular}{ |m{3cm}|m{1cm}|m{4.5cm}|m{4.5cm}| } 
		
		\hline
		\textbf{Method} & \textbf{Return Value} & \textbf{Pre-state}($S$: global state)  & \textbf{Post-state}( $S'$: future state of $S$ such that $S \sqsubset S'$)  \\ 
		
		\hline
		$\addv (u)$  & $true$ &$ S: \langle u \notin S.\abgv \rangle $ &$ S':\langle u \in S'.\abgv \rangle$	 \\
		\hline 
		$\addv (u)$  & $false$ & Never the case	 &Never the case 	 \\
		\hline 
		$\remv (u)$  & $true$ &$ S: \langle u \in S.\abgv \rangle$ &$ S':\langle u \notin S'.\abgv \rangle$	 \\
		\hline 
		$\remv (u)$  & $false$ &$ S: \langle u \notin S.\abgv \rangle$ &Same as pre-state \\
		\hline 
		$\adde (u, v)$  & $true$ &$ S: \langle u,v \in S.\abgv \wedge  ((u,v) \in S.\abge \vee (u,v) \notin S.\abge ) \rangle$ & $ S':\langle (u,v \in S'.\abgv \wedge (u,v) \in S'.\abge \rangle$	 \\
		\hline 
		$\adde(u, v)$  & $false$ &$ S: \langle  u \notin S.\abgv \vee v \notin S.\abgv \rangle$ &Same as pre-state \\
		\hline 
		$\reme(u, v)$  & $true$ &$ S: \langle u,v \in S.\abgv \wedge  ((u,v) \in S.\abge) \vee (u,v) \notin S.\abge ) \rangle$ &$ S':\langle u,v \in S'.\abgv) \wedge (u,v) \notin S'.\abge \rangle$	 \\	
		\hline 
		$\reme(u, v)$  & $false$ &$ S: \langle u\notin S.\abgv \vee v\notin S.\abgv \rangle $ & Same as pre-state \\ 
		\hline 
		$\conv(u)$  & $true$ &$ S: \langle u \in S.\abgv \rangle$ &Same as pre-state \\
		\hline 
		$\conv(u)$  & $false$ &$ S: \langle u \notin S.\abgv \rangle$ &Same as pre-state \\ 
		\hline 
		$\cone(u, v)$  & $true$ &$ S: \langle u,v \in S.\abgv \wedge  (u,v) \in S.\abge  \rangle$ & Same as pre-state \\
		\hline 
		$\cone(u, v)$  & $false$ &$ S: \langle u\notin S.\abgv \vee v\notin S.\abgv \vee  (u,v) \notin S.\abge \rangle$ &same as pre-state	 \\
		
		\hline
		
	\end{tabular}
\end{table*}
\begin{definition}
	\label{def:abgV}
	$S.\abgv \equiv \{v | (v \in \vnodes{S}) \land (S.\vh \rightarrow^* S.v) \land (\neg S.v.marked)\}$.
\end{definition}

\noindent This definition of \abgv captures the set of all vertices of $\abg$ for the global state $S$. It consists of all the \vnode{s} that are reachable from $S.\vh$ and are not marked for deletion. 

\begin{definition}
	\label{def:abgE}
	$S.\abge \equiv \{e | (e \in S.enodes) \land ((u, v) \subseteq S.\abgv) \land \\ (S.u.\eh \rightarrow^* S.e) \land (\neg S.e.marked) \land (S.e.val = S.v.val)\}$.
\end{definition}

\noindent This definition of $\abge$ captures the set of all edges of $\abg$ for the global state $S$. Informally it consists of all the $\enode$s that connects two $\vnode$s $u$, $v$ with the edge going form $u$ to $v$. 
\section{Working of Concurrent Graph Methods}
\label{sec:working-con-graph-methods}
In this section, we describe the working of the various \mth{s} on the generic \cgds. As explained earlier, we represent the graph using adjacency list representation, which is a list of list-based sets as illustrated in the Figure \ref{fig:conGraph}. 
A set is implemented as a linked list of nodes. In each list, nodes are in sorted in key order, providing an efficient way to detect when an item is absent. The next field is a reference to the next node in the list. Each list has two kinds of nodes. In addition to regular nodes that hold items in the set, we use two sentinel nodes, called head and tail and their keys are the minimum and maximum integer values.
All the fields in the structure are declared atomic. This ensures that operations on these variables happen atomically. In the context of a particular application, the node structure can be easily modified to carry useful data (like weights etc). Algorithm \ref{alg:nadd:Vertex-g}\textbf{-}\ref{contains:Edge-g} describe all the pseudo code used in our generic \cgds. For the reference of the reader, the pseudo code of lazy list-based set and \lf set is given in the Appendix \ref{sec:app-pcode}.

\noindent \textbf{Notations used in Pseudo-Code:}\\[0.2cm]
We use $\downarrow$, $\uparrow$ to denote the input and output arguments to each method respectively. The shared memory is accessed only by invoking explicit \emph{read()}, \emph{write()} and \emph{CAS} methods. The $flag$ is a local variable which returns the status of each operation.

\subsection{\emph{Update Vertex Methods} - \addv \& \remv}
\label{sec:working-con-graph-methods:add}
\noindent The $\addv(u)$ method is given in the Algorithm \ref{alg:nadd:Vertex-g}. It simply invokes the Add method of the list-based set by passing the head of the vertex list and the key to be added. 

\begin{multicols}{3}
\begin{algorithm}[H]
\captionsetup{font=tiny}
	\caption{\addv  Method: Successfully adds $GNode(key)$ to the vertex list, if it is not present earlier, else it ignores.}
	\label{alg:nadd:Vertex-g}
	\begin{algorithmic}[1]
		\tiny
		\Procedure{\addv ($key\downarrow$, $flag\uparrow$)}{}
		\State {$G.\add(G.Head\downarrow, key\downarrow, status\uparrow)$;}  
		\State{$flag \gets status$}
	  	\EndProcedure
        \algstore{addv}
	\end{algorithmic}
\end{algorithm}
\vspace{-0.6cm}
\begin{algorithm}[H]
\captionsetup{font=tiny}
	\caption{\remv Method: removes the  $GNode(key)$ from the vertex list if it is already present. Otherwise it returns $false$ }
	\label{alg:remove:Vertex-g}
	\begin{algorithmic}[1]
	\algrestore{addv}
	\tiny
		\Procedure{\remv($key\downarrow,flag\uparrow$)}{}
		\State {$G.\rem(G.Head\downarrow, key\downarrow, status \uparrow)$;}  
		\State{$flag \gets status;$}
		\State{RemoveIncomingEdges(key$\downarrow$); /*{optional method}*/}
		\EndProcedure
		\algstore{remv}
	\end{algorithmic}
\end{algorithm}
\begin{algorithm}[H]
\captionsetup{font=tiny}
	\caption{RemoveIncomingEdge Method: This method helps remove all the incoming edges of a deleted vertex $GNode(key)$ from the graph.}
	\label{alg:HelpRemoveIncomingEdgex}
	\tiny
	\begin{algorithmic}[1]
	\algrestore{remv}
		\Procedure{RemoveIncomingEdges($key\downarrow$)}{}
		 \State{$node \gets G.Head$}
		\While{(node != G.Tail))}
		\State{$G.\rem(node.enext\downarrow, key\downarrow, flag\uparrow)$;}
		\State{$node \gets node.vnext$;}
		\EndWhile
		
		\EndProcedure
		\algstore{remie}
	\end{algorithmic}
\end{algorithm}

\vspace{-0.6cm}
\end{multicols}

\noindent When a thread wants to delete a vertex from the concurrent graph, it invokes the Remove method of the list-based set by passing the head of the vertex list and the key of the vertex to be deleted. This is described in Algorithm \ref{alg:remove:Vertex-g}.
Once a vertex has been deleted from the vertex list, its outgoing edges are logically removed automatically. This is because any operation in the edge list of the deleted vertex, will first verify for the presence of the vertex in the graph. After the deletion of the vertex in the vertex list, we must also delete the incoming edges to the deleted vertex. This is described in Algorithm \ref{alg:HelpRemoveIncomingEdgex}. The RemoveIncomingEdges method performs a traversal of the entire \vlist, to check if any of the existing reachable vertices contains an edge node corresponding to the deleted vertex in their edge list. If such an edge node is present, it is simply deleted from its edge list.\\[0.1cm] 
It is to be noted that performing the deletion of incoming edges of deleted vertices is an optional step as this does not affect the correctness of the algorithm. In other words, even if edge nodes corresponding to the deleted vertices are still reachable, no other method's correctness is affected by their presence. In later section, we present results of algorithms without removing their incoming edges.
\vspace{-3mm}
\subsection{\emph{Update Edge Methods} - \adde \& \reme }
\label{sec:working-con-graph-methods:remove}

\begin{multicols}{2}
\begin{algorithm}[H]
\captionsetup{font=tiny}
	\caption{\adde Method: $GNode(key_2)$ gets added to the edge list of $GNode(key_1).enext$, if it is not present and returns true.}
	\label{alg:add:Edge-g}
	\begin{algorithmic}[1]
	\algrestore{remie}
	\tiny
		\Procedure{\adde ($key_1\downarrow$,$key_2\downarrow, flag\uparrow$)}{}
		\State {$G.\con(G.Head\downarrow, key_1\downarrow, u\uparrow,  status1\uparrow);$}  \label{lin:adde-validate-u}
		\State {$G.\con(G.Head\downarrow, key_2\downarrow, v\uparrow status2\uparrow);$} \label{lin:adde-validate-v}
		\If {( status1 = false $\vee$  status2 = false)}
		\State{$flag \gets false$;}
        \State {$return$; }
        \EndIf
        \State {$G.\con(G.Head\downarrow, key_1\downarrow, u\uparrow, status1\uparrow);$} \label{lin:adde-validate-u-again}  
      	\If {($status1 = false $)} 
      		\State{$flag \gets false$;}
        \State {$return$; }
        \EndIf
	    \State{$G. \add(u.enext\downarrow, key_2\downarrow, status \uparrow)$;}\label{lin:adde-uv} 
	    	\State{$flag \gets status$;}
        \EndProcedure
		\algstore{adde}
	\end{algorithmic}
\end{algorithm}
\vspace{-0.6cm}
\begin{algorithm}[H]
\captionsetup{font=tiny}
	\caption{\reme Method: $GNode(key_2)$ gets removed from the edge list of $GNode(key_1).enext$, if it is present. Returns unsuccessful if the edge is not present earlier.}
	\label{alg:remove:Edge-g}
	\begin{algorithmic}[1]
	\algrestore{adde}
	\tiny
		\Procedure{\reme ($key_1\downarrow$, $key_2\downarrow, flag\uparrow$)}{}
    	\State {$G.\con(G.Head\downarrow, key_1\downarrow, u\uparrow,  status1\uparrow);$}   \label{lin:reme-validate-u}
		\State {$G.\con(G.Head\downarrow, key_2\downarrow, v\uparrow, status2\uparrow);$}\label{lin:reme-validate-v}
		\If {( status1 = false $\vee$  status2 = false)}
		\State{$flag \gets false$;}
        \State {$return$; }
        \EndIf
        \State {$G.\con(G.Head\downarrow, key_1\downarrow, u\uparrow,  status1\uparrow);$}   \label{lin:reme-validate-u-again}
      	\If {($status1 = false $)} 
      		\State{$flag \gets false$;}
        \State {$return$; }
        \EndIf
	    \State{$G.\rem(u.enext\downarrow, key_2\downarrow, status \uparrow)$;}\label{lin:reme-uv}
	    \State{$flag \gets status$;}
	    \EndProcedure
        \algstore{reme}
	\end{algorithmic}
\end{algorithm}
\end{multicols}
\noindent The $\adde(u, v)$ method starts by checking for the presence of vertices $u$ and $v$ in the vertex list of the graph by invoking the $\con$ method in the Lines \ref{lin:adde-validate-u} \& \ref{lin:adde-validate-v} respectively. After this, once again $u$ is validated in the Line \ref{lin:adde-validate-u-again}, the reason for this is explained by an example in Figure \ref{fig:noseqhist}.

\noindent
\begin{minipage}{\linewidth}
      \centering
\begin{minipage}{.466\textwidth}
  After successful checking for the presence of both the vertices $\adde$ invoked the $\add(v)$ in the $u$'s adjacency list($enext$) of respective \ds in the Line \ref{lin:adde-uv}. It adds the edge $(u,v)$, if not present earlier and then it returns $true$. If the edge is already in the graph it simply returns $true$. But if either the vertices $u$ or $v$ is not present, it returns $false$. The generic $\adde$ method is given in the Algorithm \ref{alg:add:Edge-g}.\\
\noindent The $\reme(u,v)$ method proceeds similar to the $\adde(u,v)$, it first checks for the presence of vertices $u$ and $v$ in the vertex list of the graph by invoking the $\con$ method in the Lines \ref{lin:reme-validate-u} \& \ref{lin:reme-validate-u} respectively. After this, once again $u$ is checked for presence in the Line \ref{lin:reme-validate-u-again} for the same reason as described earlier. After successful checking for the presence of both the vertices, $\reme$ invokes the $\rem(v)$ in the $u$'s adjacency list($enext$). It removes the edge $(u,v)$, if it is present in the graph  and then it returns $true$. But if either the vertices $u$ or $v$ are not present, it returns $false$. The generic $\reme$ method is given in the Algorithm \ref{alg:remove:Edge-g}.
\end{minipage}
\hspace{0.05\linewidth}
\begin{minipage}{.466\textwidth}
\begin{figure}[H]
	\centerline{\scalebox{0.33}{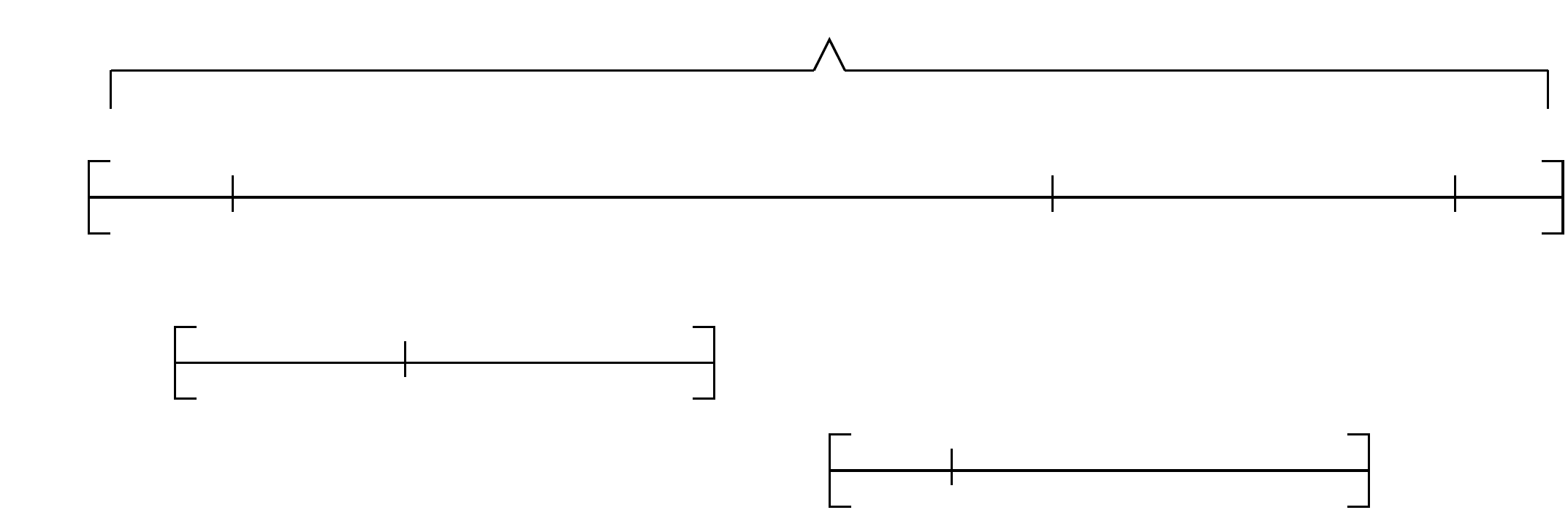}}
	\caption{This figure depicts why we need an additional check to locate vertex in AddEdge. A thread $T_1$ tries to perform \adde(u, v, true), first invokes $contains(u)$. Just after $T_1$ has verified vertex $u$ to be present in vertex list, thread $T_2$ deletes vertex $u$. Also vertex $v$ gets added by thread $T_3$ just before $T_1$ verifies it. Now thread $T_1$ has successfully tested for the presence of vertices $u$ and $v$ in the vertex list, and then it proceeds to add edge $(u,v)$, returning true. However, as is evident, in no possible sequential history equivalent to the given concurrent execution will both the vertices u and v exist together. Hence an additional check must be performed before proceeding to actually add the edge. With this additional check, in this scenario, AddEdge(u,v) will return false on checking that vertex u has been deleted.}
	\label{fig:noseqhist}
\end{figure}
 \end{minipage}
\end{minipage}
\noindent

\subsection{\emph{Read-Only Methods} - \conv \& \cone}
\noindent Method $\conv(u)$ simply invokes the Contains method of the list-based set by passing the head of the vertex list and the vertex key to be searched. The $\cone(u,v)$ method first invokes the Contains method of the list-based set for each of the vertex keys. If they are found in the vertex list, it then calls for Contains in the edge list of the first vertex. These methods return $true$ if the vertex/edge node it was searching for is present and unmarked, otherwise returns $false$. The generic $\conv$ \& $\cone$ methods are given in the Algorithm \ref{contains:Vertex-g} \& \ref{contains:Edge-g} respectively. These read-only methods are wait-free based on our assumption that the $\con$ method of the underlying list is wait-free.
\begin{multicols}{2}
\begin{algorithm}[H]
\captionsetup{font=tiny}
	\caption{\conv Method: Returns $true$ if $GNode(key)$ is present in vertex list and returns $false$ otherwise.}
	\label{contains:Vertex-g}
	\begin{algorithmic}[1]
	\algrestore{reme}
	\tiny
		\Procedure{\conv ($key\downarrow, flag\uparrow$)}{}
		\State {$G.\con(G.Head\downarrow, key\downarrow, u\uparrow, status \uparrow)$;}  
		\State{$flag \gets status;$}
		\EndProcedure
		\algstore{conv}
	\end{algorithmic}
\end{algorithm}
\vspace{-0.6cm}
\begin{algorithm}[H]
\captionsetup{font=tiny}
	\caption{\cone Method: Returns $true$ if $GNode(key_2)$ is part of the edge list of $GNode(key_1)$ and returns $false$ otherwise.}
	\label{contains:Edge-g}
	\begin{algorithmic}[1]
	\algrestore{conv}
	\tiny
		\Procedure{\cone ($key_1\downarrow, key_2\downarrow, flag\uparrow$)}{}
     	\State {$G.\con(G.Head\downarrow, key_1\downarrow, u\uparrow,  status1\uparrow);$}   
		\State {$G.\con(G.Head\downarrow, key_2\downarrow, v\uparrow, status2\uparrow);$}
		\If {( status1 = false $\vee$  status2 = false)}
		\State{$flag \gets false$;}
        \State {$return$; }
        \EndIf
        \State{$G.\con(u.enext\downarrow, key_2\downarrow, x\uparrow, status\uparrow)$;}
        \State{$flag \gets status$;}
		\EndProcedure
	\end{algorithmic}
\end{algorithm}
\vspace{-0.6cm}
\end{multicols}
\vspace{-3mm}
\subsection{Correctness: Linearization Points}
\noindent  In this subsection, we describe the \textit{Linearization Points}(LP) \cite{HerlWing:1990:TPLS} of all methods of our \cgds. Due to lack of space, the proof of the correctness of \cgds is given in the full paper\cite{PeriSS16}. We try to formalise the proof of our \cgds based on $\lp$ events of the methods and is also based on \cite{Peri+:LPs:TR:2017}.\\[0.2cm]
The $\lp$ of $\addv(key, true)$, $\remv(key, true)$, $\remv(key, false)$, \\ $\wconv(key, true)$ \& $\wconv(key, false)$ is the linearization point of $\add(key, true)$, $\rem(key, true)$, $\rem(key, false)$, $\con(key, true)$ \& $\con(key, false)$ in the corresponding concurrent list-based set implementation.\\[0.2cm]
 \noindent We linearize a successful $\adde(key_1, key_2, true)$ method call within its execution interval: (1) if there is no successful concurrent delete vertex on $key_1$ and $key_2$, the $LP$ is defined as the LP of the $\add(key, true)$; (2) if there is a successful concurrent delete vertex on $key_1$ or $key_2$ or both, the $LP$ is the point immediately before the first $LP$ of successful concurrent delete on vertex $key_1$ or $key_2$.\\[0.2cm]
 In case of an unsuccessful $\adde(key_1, key_2, false)$ method call, the $LP$ is defined to be within its execution interval: (1) if there is no successful concurrent add vertex on $key_1$ and $key_2$, $LP$ is defined as the LP of the $\con(key, false)$ when there is no concurrent Add method; (2) if there is a successful concurrent add vertex on $key_1$ or $key_2$ or both, it is linearized at the point immediately before the $LP$ of the first successful concurrent add vertex on $key_1$ or $key_2$. \\[0.2cm]
We linearize a successful $\reme$ method call within its execution interval: (1) if there is no successful concurrent delete vertex, the $LP$ is defined as the LP of the $\rem(key, true)$; (2) if there is a successful concurrent delete vertex on $key_1$ or $key_2$ or both, it is linearized just before the $LP$ of the first successful concurrent delete vertex on $key_1$ or $key_2$. In case of an unsuccessful $\reme(key_1, key_2, false)$ method call, the $LP$ is defined to be the same as the LP of $\adde(key_1, key_2, false)$ within its execution interval.\\[0.2cm]
We linearize a successful $\cone(key_1, key_2)$ method call within its execution interval: (1) if there is no successful concurrent delete vertex on $key_1$ and $key_2$, the $LP$ is defined as the LP of $\con(key, true)$; (2) if there is a successful concurrent delete vertex on $key_1$ or $key_2$, it is linearized immediately before the $LP$ of the first successful concurrent delete on the corresponding vertex. In case of an unsuccessful $\wcone(key_1, key_2, false)$ method call, the $LP$ is defined to be within its execution interval: (1) if there is no successful concurrent add edge $(key_1, key_2)$, the $LP$ is the LP of the $\con(key, false)$; (2) if there is a successful concurrent add edge on $(key_1, key_2)$, it is linearized immediately before the $LP$ of that successful concurrent add edge. 

 \begin{lemma}
\label{lem:linealizability-g}
The history $H$ generated by the interleaving of any of the methods of the \cgds, is \lble. \\
\textnormal{Proof in \cite{PeriSS16}}.
\end{lemma}
\begin{lemma}
\label{lem:app-prog:add:rem-g}
The methods of the concurrent graph data structure have the same progress guarantees as provided by the concurrent list-based set implementation used to implement it.\\
\textnormal{Proof in \cite{PeriSS16}}.
\end{lemma}

\vspace{-7mm}
\section{Simulation Results \& Analysis}
\label{sec:simulation-results}

\vspace{-2mm}
\begin{figure*}[!hbtp]
	\captionsetup{font=scriptsize}
	\begin{subfigure}[b]{0.3\textwidth}
		\setlength{\belowcaptionskip}{-2mm}
		\captionsetup{font=scriptsize, justification=centering}
		\caption{CV:15\%, CE:15\%, AddE:25\%, DelE:10\%, AddV:25\% \& DelV:10\%.}
		\centering
		\resizebox{\linewidth}{!}{
			\begin{tikzpicture} [scale=0.30]
			\begin{axis}[legend style={at={(0.5,1)},anchor=north},
			xlabel=No of threads,
			ylabel=Throughput ops/sec, ylabel near ticks]
			\addplot table [x=Thread, y=$Sequential$]{results/gdsdist70301000.dat};
			\addplot table [x=Thread, y=$Coarse$]{results/gdsdist70301000.dat};
			\addplot table [x=Thread, y=$HoH$]{results/gdsdist70301000.dat};
			\addplot table [x=Thread, y=$Lazy$]{results/gdsdist70301000.dat};
			\addplot table [x=Thread, y=$Lock-Free$]{results/gdsdist70301000.dat};
			\end{axis}
			\end{tikzpicture}
		}
		\label{fig:subfig-70-30}
	\end{subfigure}
	\begin{subfigure}[b]{0.3\textwidth}
		\setlength{\belowcaptionskip}{-2mm}
		\captionsetup{font=scriptsize, justification=centering}
		\caption{CV:40\%, CE:40\%, AddE:7\%, DelE:3\%, AddV:7\% \& DelV:3\%}   
		\centering
		\resizebox{\linewidth}{!}{
			\begin{tikzpicture}[scale=0.30]
			\begin{axis}[
			xlabel=No of threads,
			ylabel=Throughput ops/sec, ylabel near ticks]
			\addplot table [x=Thread, y=$Sequential$]{results/gdsdist20801000.dat};
			\addplot table [x=Thread, y=$Coarse$]{results/gdsdist20801000.dat};
			\addplot table [x=Thread, y=$HoH$]{results/gdsdist20801000.dat};
			\addplot table [x=Thread, y=$Lazy$]{results/gdsdist20801000.dat};
			\addplot table [x=Thread, y=$Lock-Free$]{results/gdsdist20801000.dat};
			\end{axis}
			\end{tikzpicture}
		}
		\label{fig:subfig-40-60}
	\end{subfigure}
	\begin{subfigure}[b]{0.3\textwidth}
		\captionsetup{font=scriptsize, justification=centering}
		\setlength{\belowcaptionskip}{-2mm}
		\caption{AddE:50\%, DelE: 50\% and rest are 0\%}
		\centering
		\resizebox{\linewidth}{!}{
			\begin{tikzpicture}[scale=0.30]
			\begin{axis}[legend style={at={(0.5,0.65)},anchor=north},
			xlabel=No of threads,
			ylabel=Throughput ops/sec, ylabel near ticks]
			\addplot table [x=Thread, y=$Sequential$]{results/gdsdist50501000.dat};
			\addlegendentry{$Sequential$}
			\addplot table [x=Thread, y=$Coarse$]{results/gdsdist50501000.dat};
			\addlegendentry{$Coarse$}
			\addplot table [x=Thread, y=$HoH$]{results/gdsdist50501000.dat};
			\addlegendentry{$HoH$}
			\addplot table [x=Thread, y=$Lazy$]{results/gdsdist50501000.dat};
			\addlegendentry{$Lazy$}
			\addplot table [x=Thread, y=$Lock-Free$]{results/gdsdist50501000.dat};
			\addlegendentry{$Lock\text{-}Free$}
			\end{axis}
			\end{tikzpicture}
		}
		\label{fig:subfig10}
	\end{subfigure}
	\vspace{-4mm}
	\setlength{\belowcaptionskip}{-2mm}
	\caption{Generic Graph Data-Structure Results} 
	\label{fig:subfig1.a.4-ds}
\end{figure*}
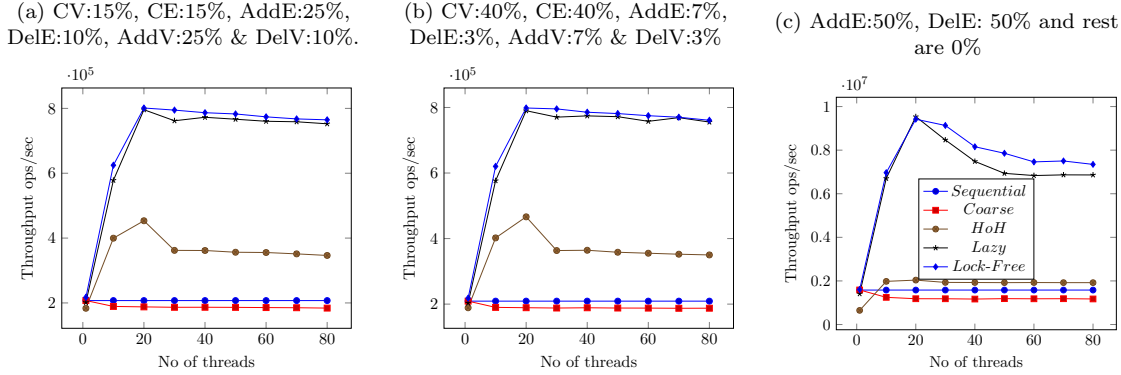

\noindent We performed our tests on 2 sockets, 10 cores per socket, Intel Xeon (R) CPU E5-2630 v4 running at 2.20 GHz frequency. Each core supports 2 hardware threads. Every core's L1 has 64k, L2 has 256k cache memory are private to that core; L3 cache (25MB) is shared across all cores of a processors. The tests were performed in a controlled environment, where we were the sole users of the system. The implementation\footnote{The complete source code is available on Github \cite{Pdcrl-ConcurrentGraphDS2017}.} has been done in C++ (without any garbage collection) and threading is achieved by using Posix threads and all the programs were optimized at O3 level. 

In the experiments conducted, we start with an initial complete graph. When the program starts, it creates fixed number of threads and each thread randomly performs a set of operations chosen by a particular workload distribution. Here, the evaluation metric used is the number of operations completed in unit time. We measure throughput obtained on running the experiment for 20 seconds and present the results for the following workload distributions: (1) \textit{Update-dominated}: $25\%$ $\addv$, $25\%$ $\adde$, $10\%$  $\remv$, $10\%$  $\reme$, $15\%$ $\conv$ and $15\%$  $\cone$; (2) \textit{Contains-dominated}: $40\%$ $\conv$, $40\%$  $\cone$, $7\%$ $\addv$, $7\%$ $\adde$, $3\%$ $\remv$ and $3\%$ $\reme$; (3) \textit{Edge-updates}: $50\%$ $\adde$, $50\%$ $\reme$ and rest are $0\%$. Figure \ref{fig:subfig1.a.4-ds} depicts the results for the data-structure methods. Each data point is obtained after averaging for 5 iterations. We assume that duplicate vertices are not inserted.\\

\noindent
It is to be noted that all the variants of the concurrent data-structure are implemented without Deletion of Incoming Edges (DIE) for deleted vertices since it achieves higher throughput than the one with DIE. This can be attributed to the observation that it is cost inefficient to traverse all the vertices to search for the incoming edges of the deleted vertices. 
We tested different variants of the data-structure for different number of threads - LockFree implementation \cite{Harrisdisc01}, Lazy list-based set implementation\cite{Heller-PPL2007}, hand-over-hand coupling list, CoarseLock\cite[Chap 9]{MauriceNir}: which supports concurrent operations by acquiring a global lock and the sequential implementation. The figures depict that the performance of the concurrent data structure is similar to that of the lazy list-based set performance. We noted on an average $5$x increased throughput. 
\vspace{-3mm}
\section{Conclusion \& Future Direction}
\label{sec:conclusion-future-directions}
In this paper, we have shown how to construct a fully dynamic \cgds, which allows threads to concurrently add/delete vertices/edges. The graph is constructed by the composition of the well known concurrent list-based set data structure from the literature. Our construction is generic, in the sense that it can be used to obtain various progress guarantees, depending on the granularity of the underlying concurrent set implementation - either blocking or non-blocking. We believe that there are many applications that can benefit from this concurrent graph structure. An important application that inspired us is \sgt in Databases and Transactional Memory. For proving \lbty, we have identified the linearization points of all the methods. We have compared the performance of the different variants of concurrent data-structure and we achieve on an average \emph{5}x increased throughput.

\bibliographystyle{plain}
\bibliography{biblio}

\newpage

\appendix

\begin{center}
\Large\textbf{Appendix}
\end{center}
\label{sec:appendix-cgds}
\vspace{-3mm}
\section{Correctness:Linearization Points}
\label{sec:app-corectness-lp}
In this subsection, we define the \textit{Linearization Points (LPs)} based on \cite{HerlWing:1990:TPLS} of all methods of our \cgds and the proof technique is based on \cite{Peri+:LPs:TR:2017}. To show that a \cgds
is a linearizable implementation we need to 
identify the correct linearization point. 
\subsection{Linearization Points:based on \fg graph }
\noindent Here, we list the linearization points (\lp{s}) of each method. Note that each method of the list can return either $true$ or $false$ except \naddv. So, we define the $\lp$ bellow:
\begin{enumerate}
\item $\addv(key, true)$:  $write(n_1.next, n_3)$ in Line \ref{lin:add6} of $\add$ method. which implies that the key $u$ is not present and the effect of this method actually happens at this line, such that a new vertex node is now made reachable from $\vh$ of the vertex list. It can be seen that $\addv$ never returns false which follows from the \sspec.
\item $\remv(key, true)$: $write(n_2.marked, true)$ in Line \ref{lin:rem4} of $\rem$ method, this means the key $u$ is already present in the vertex list.
\item $\remv(key, false)$: $read(n_2.val)$ in Line \ref{lin:rem3} of $\rem$ method, where the $key$ is found to be not present in the vertex list. where the $key$ is not present in the vertex list. 
\item $\wconv(key, true)$: $read(n.marked)$ in \lineref{cont-tf} of $\con$ method \ref{step:contT}, where the $key$ is unmarked in the vertex list.
\item $\wconv(key, false)$: $\lp$ is the last among the following lines executed. There are three cases here: 
\begin{itemize}
	\item [(a)] $read(n.val) \neq key$ in \lineref{cont-tf} of $\con$ method is the $\lp$, in case of no concurrent $\add(key, true)$.
	\item [(b)] $read(n.marked)$ in \lineref{cont-tf} of $\con$ method is the $\lp$, in case of no concurrent $\add(key, true)$ (like the case of \stref{contT}).
    \item [(c)] in case of concurrent $\add(key, true)$ by another thread, we add a dummy event just before Line \ref{lin:add6} of $add(key, true)$. This dummy event is the $\lp$ of $\con$ method if: (i) if in the post-state of $read(n.val)$ event in \lineref{cont-tf} of \con method, $n.val \neq key$ and $write(n_1.next, n_3)$ (with $n_3.val = key$) in Line \ref{lin:add6} of \add method executes before this $read(n.val)$. (ii) if in the post-state of $read(n.marked)$ event in \lineref{cont-tf} of \con method, $n.marked = true$ and $write(n_1.next, n_3)$ (with $n_3.val = key$) in Line \ref{lin:add6} of \add method executes before this $read(n.marked)$.
    \end{itemize}
\item $\adde(key_1, key_2, true)$: We linearize a successful $\adde(key_1, key_2)$ method call within its execution interval at the earlier of the following points; (1) if there is no successful concurrent delete vertex on $key_1$ and $key_2$, the $LP$ is defined as the last of $read(key_2)$  and $write(key_1.enext, key_3)$ in the \add method, depending upon the execution. (2) if there is a successful concurrent delete vertex on $key_1$ or $key_2$ or both, the $LP$ is the point immediately before the first $LP$ of successful concurrent delete on vertex $key_1$ or $key_2$.
\item $\adde(key_1, key_2, false)$:  The $LP$ is defined to be within its execution interval at the earlier of the following points; (1) if there is no successful concurrent add vertex on $key_1$ and $key_2$, $LP$ is the last of $read(key_1)$/$read(key_2)$ and $read(key_1.marked)$/$read(key_2.marked)$ in the \con methods depending upon the execution. (2) if there is a successful concurrent add vertex on $key_1$ or $key_2$ or both, it is linearized at the point immediately before the $LP$ of the first successful concurrent add vertex on $key_1$ or $key_2$. 
\item $\reme(key_1, key_2, true)$: We linearize a successful $\reme$ method call within its execution interval at the earlier of the following points; (1) if there is no successful concurrent delete vertex, the $LP$ is later of $read(key_2.val)$ and $write(key_2.marked, true)$ in \rem method, based on the execution. If the edge $(key_1, key_2)$ is already present in the edge list of the vertex $key_1$, the $\lp$ is the logically marked as deleted. If the edge $(key_1, key_2)$ is not present in the edge list of the vertex $key_1$, the $\lp$ is the $read(key_2)$ in the \rem method. (2) if there is a successful concurrent delete vertex on $key_1$ or $key_2$ or both, it is linearized just before the $LP$ of the first successful concurrent delete vertex on $key_1$ or $key_2$.
\item $\reme(key_1, key_2, false)$:   The $LP$ is defined to be within its execution interval at the earlier of the following points; The $LP$s remain same as the $LP$s of the $\adde(key_1, key_2)$ returning $false$.
\item $\wcone(key_1, key_2, true)$: We linearize a successful $\cone(key_1, key_2)$ method call within its execution interval at the earlier of the following points: (1) if there is no successful concurrent delete vertex on $key_1$ and $key_2$, the $LP$ is $read(n.marked)$ in \lineref{cont-tf} of $\con$ method \label{step:contT}, where the $key_1/key_2$ is unmarked in the vertex list. (2) if there is a successful concurrent delete vertex on $key_1$ or $key_2$, it is linearized immediately before the $LP$ of the first successful concurrent delete on corresponding vertex.

\item $\wcone(key_1, key_2, false)$: The $LP$ is defined to be within its execution interval at the earlier of the following points;  (1) if there is no successful concurrent add edge $(key_1, key_2)$, the $LP$ is, (a) last of $read(key_1)$ or $read(key_1.marked)$ in Line \con method, (b) last of $read(key_2)$ or $read(key_2.marked)$ in Line \con method, depending upon the execution. (2) if there is a successful concurrent add edge on $(key_1, key_2)$, it is linearized immediately before the $LP$ of that successful concurrent add edge. 

\end{enumerate}

\subsection{Linearization Points:based on \lf graph }
\noindent Here, we list the \lp{s} of each method. As each method of the list can return either $true$ or $false$ except \naddv. So, we define the $\lp$ based on that and given bellow:

\begin{enumerate}
\item $\lfaddv(key, true)$:  Successful CAS in Line \ref{lin:naddedv} of $\lfadd$ method.
\item $\lfremv(key, true)$:  Successful CAS for logical marked in the \lineref{lf-logical-marked} of $\lfrem$ method.
\item $\lfremv(key, false)$: The LP be in the \lineref{lf-readn1} or \lineref{lf-readn1-next} in the $\lfloct$ method.
\item $\wfconv(key, true)$: $read(v.vnext);$ in the \lineref{con-lf-lp} of $\wfcon$ method.
\item $\wfconv(key, false)$: $\lp$ is the last among the following lines executed. There are two cases depending on concurrent $\lfaddv$ method: 
\begin{itemize}
	\item [(a)] with no successful concurrent $\lfaddv(key, true)$:  $read(v.vnext);$ in the \lineref{con-lf-lp} of $\wfcon$ method.
	\item [(b)] with successful concurrent $\lfaddv(key, true)$: we add a dummy event just before \lineref{naddedv} of \lfaddv method. This dummy event is the \emph{\lp} of the \wfconv method.
    \end{itemize}
    
\item $\lfadde(key_1, key_2, true)$: We linearize a successful $\lfadde(key_1, key_2)$ method call within its execution interval at the earlier of the following points; (1) if there is no successful concurrent \lfremv on $key_1$ and $key_2$, the $LP$ is the successful CAS in Line \ref{lin:naddedv} of $\lfadd$ method if $(read(n_2.val) \neq key$) otherwise the $\lp$ be in the \lineref{lf-readn1} or \lineref{lf-readn1-next} in the $\lfloct$ method. (2) if there is a successful concurrent \lfremv on $key_1$ or $key_2$ or both, the $\lp$ is the point immediately before the first $\lp$ of successful concurrent delete on vertex $key_1$ or $key_2$.
\item $\lfadde(key_1, key_2, false)$:   We linearize an unsuccessful $\lfadde(key_1, key_2)$ method call within its execution interval at the earlier of the following points; (1) if there is no successful concurrent \lfaddv on $key_1$ and $key_2$, the $LP$ be in the \lineref{lf-readn1} or \lineref{lf-readn1-next} in the $\lfloct$ method if $(read(n_2.val) = key$). (2) if there is a successful concurrent \lfaddv on $key_1$ or $key_2$ or both, the $\lp$ is the point immediately before the first $\lp$ of successful concurrent add on vertex $key_1$ or $key_2$.
\item $\lfreme(key_1, key_2, true)$:  We linearize a successful $\lfreme(key_1, key_2)$ method call within its execution interval at the earlier of the following points; (1) if there is no successful concurrent \lfremv on $key_1$ and $key_2$, the $LP$ be in the \lineref{lf-logical-marked} logical marked in the \lfremv method if $(read(n_2.val) = key$) otherwise \lineref{lf-readn1} or \lineref{lf-readn1-next} in the $\lfloct$ method . (2) if there is a successful concurrent \lfremv on $key_1$ or $key_2$ or both, the $\lp$ is the point immediately before the first $\lp$ of successful concurrent delete on vertex $key_1$ or $key_2$.
\item $\lfreme(key_1, key_2, false)$: We linearize an unsuccessful $\nreme(key_1, key_2)$ method call within its execution interval at the earlier of the following points; (1) if there is no successful concurrent \lfaddv on $key_1$ and $key_2$, the $LP$ be in the \lineref{lf-readn1} or \lineref{lf-readn1-next} in the $\lfloct$ method . (2) if there is a successful concurrent \lfaddv on $key_1$ or $key_2$ or both, the $\lp$ is the point immediately before the first $\lp$ of successful concurrent add on vertex $key_1$ or $key_2$.

\item $\wfcone(key_1, key_2, true)$: We linearize a successful $\wfcone(key_1, key_2)$ method call within its execution interval at the earlier of the following points; (1) if there is no successful concurrent \lfremv on $key_1$ and $key_2$, the $\lp$ be the last of these execution $read(key)$  or $read(key_1.marked)$ in the \lineref{conv6-lf} of $\wfcon$ method. (2) if there is a successful concurrent \lfremv on $key_1$ or $key_2$ or both, the $\lp$ is the point immediately before the first $\lp$ of successful concurrent delete on vertex $key_1$ or $key_2$.
\item $\wfcone(key_1, key_2, false)$:  We linearize an unsuccessful $\wfcone(key_1, key_2)$ method call within its execution interval at the earlier of the following points; (1) if there is no successful concurrent \lfadde on $(key_1, key_2)$, the $\lp$ be the last of these execution $read(key)$  or $read(key_1.marked)$ in the \lineref{conv6-lf} of $\wfcon$ method. (2) if there is a successful concurrent \lfadde on $(key_1, key_2)$, the $\lp$ is the point immediately before the $\lp$ of successful concurrent \lfadde$(key_1, key_2)$.
\end{enumerate}


\section{PCode}
\label{sec:app-pcode}
\subsection{\hoh PCode}
\begin{multicols}{2}
	\tiny
	\begin{algorithm}[H]
		\captionsetup{font=scriptsize}
		\caption{\hloct Method: Takes $key$ as input and returns the corresponding pair of neighboring $\node$ $\langle n_1, n_2 \rangle$. Initially $n_1$ and $n_2$ are set to $null$.}
		\label{alg:hlocate}
		\begin{algorithmic}[1]
			\scriptsize
			\Procedure{\hloct ($key\downarrow, n_1\uparrow, n_2\uparrow$)}{}
			\State {$lock.acquire(\hhead)$;} \label{lin:hloc2}
			\State{$\hnode$ $n_1 = \hhead;$} \label{lin:hloc3}
			\State{$\hnode$ $n_2 = n_1.next;$}\label{lin:hloc4}
			\State {$lock.acquire(n_2)$;}  \label{lin:hloc5}
			\While{($read(n_2.val) < key)$} \label{lin:hloc6}
			\State {$lock.release(n_1)$;} \label{lin:hloc7}
			\State{$ n_1 \gets n_2;$} \label{lin:hloc8}
			\State{$n_2 \gets n_2.next$} \label{lin:hloc9}
			\State {$lock.acquire(n_2)$;}  \label{lin:hloc10}
			\EndWhile\label{lin:hloc11}
			\EndProcedure
			\algstore{hlocate}
		\end{algorithmic}
	\end{algorithm}
	\begin{algorithm}[H]
		\captionsetup{font=scriptsize}
		\caption{\hcon Method: Returns $true$ if $key$ is part of the set and returns $false$ otherwise.}
		\label{alg:hcontains}
		\scriptsize
		\begin{algorithmic}[1]
			\algrestore{hlocate}
			\Procedure{\hcon ($key\downarrow, flag\uparrow$)}{}
			\State{$\hloct(key\downarrow, n_1 \uparrow, n_2\uparrow)$;}\label{lin:hcon2} 
			\If {$(read(n_2.val) = key) $}  \label{lin:hcon3}
			\State {$flag$ $\gets$ $true$;}\label{lin:hcon4}
			\Else
			\State {$flag$ $\gets$ $false$;} \label{lin:hcon5}
			\EndIf
			\State {$lock.release(n_1)$;}
			\State {$lock.release(n_2)$;}
			\State {$return$;}
			\EndProcedure
			\algstore{hcon}
		\end{algorithmic}
	\end{algorithm}
	
	\begin{algorithm}[H]
		\captionsetup{font=scriptsize}
		\caption{\hadd Method: $key$ gets added to the list if it is not already part of the list. Returns $true$ on successful add and returns $false$ otherwise.	}
		\label{alg:hadd}
		\begin{algorithmic}[1]
			\scriptsize
			\algrestore{hcon}
			\Procedure{\hadd ($key\downarrow, flag \uparrow$)}{}
			\State{$\hloct(key\downarrow, n_1 \uparrow, n_2\uparrow)$;}   \label{lin:hadd2}  
			\If {$(read(n_2.val) \neq key$)} \label{lin:hadd3}
			\State {$write(n_3, \text{new \hnode}(key))$;} \label{lin:hadd4}
			\State {$write(n_3.next, n_2)$;} \label{lin:hadd5}
			\State {$write(n_1.next, n_3)$;} \label{lin:hadd6}
			\State {$flag$ $\gets$ $true$;}
			\Else
			\State {$flag$ $\gets$ $true$;}    
			\EndIf
			\State {$lock.release(n_1)$;}
			\State {$lock.release(n_2)$;}
			\State{$return$;}
			\EndProcedure
			\algstore{hadd}
		\end{algorithmic}
	\end{algorithm}
	\begin{algorithm}[H]
		\captionsetup{font=scriptsize}
		\caption{\hrem Method: $key$ gets removed from the list if it is already part of the list. Returns $true$ on successful remove otherwise returns $false$.}
		\label{alg:hremove}
		\begin{algorithmic}[1]
			\scriptsize
			\algrestore{hadd}
			\Procedure{\hrem ($key\downarrow, flag\uparrow$)}{}
			\State{$\hloct(key\downarrow, n_1 \uparrow, n_2\uparrow)$;}\label{lin:hrem2} 
			\If {$(read(n_2.val) = key)$} \label{lin:hrem3} 
			\State {$write(n_1.next, n_2.next)$;} \label{lin:hrem5}
			\State {$flag$ $\gets$ $true$;}\label{lin:hrem8}
			\Else
			\State {$flag$ $\gets$ $false$;}
			\EndIf
			\State {$lock.release(n_1)$;} \label{lin:hrem6}
			\State {$lock.release(n_2)$;}
			\State {$return$;}
			\EndProcedure
		\end{algorithmic}
	\end{algorithm}
\end{multicols}
\subsection*{\textbf{The LPs of the \hoh}}
\label{subsec:lps-hoh}
\noindent \\ Here, we list the linearization points (\lp{s}) of each method of \hoh. Each method of the list can return either $true$ or $false$. So, we define the $\lp$ for six methods:

\begin{enumerate}
	\item $\hadd(key, true)$: $write(n_1.next, n_3)$ in Line \ref{lin:hadd6} of $\hadd$ method.
	\item $\hrem(key, true)$: $write(n_1.next, n_2.next)$ in Line \ref{lin:hrem5} of $\hrem$ method.
	\item $\hrem(key, false)$: $(read(n_2.val))$ in Line \ref{lin:hrem3} of $\hrem$ method.
	\item $\hcon(key, true)$: $read(n.val)$ in Line \ref{lin:hcon3} of $\hcon$ method.
	\item $\hcon(key, false)$:$read(n.val)$ in Line \ref{lin:hcon3} of  $\hcon$ method.
\end{enumerate}
\subsection{Lazy-List PCode}
\begin{multicols}{3}
	\tiny
	\begin{algorithm}[H]
		\captionsetup{font=scriptsize}
		\caption{\valid Method: Takes two nodes, $n_1, n_2$, each of type $\node$ as input and validates for presence of nodes in the list and returns $true$ or $false$}
		\label{alg:validate}
		\begin{algorithmic}[1]
			\scriptsize
			\Procedure{\valid($n_1 \downarrow, n_2\downarrow, flag \uparrow$)}{}
			\If {($read(n_1.marked) = false) \wedge (read(n_2.marked) = false) \wedge (read(n_1.next) = n_2$) )}
			\State {$flag$ $\gets$ $true$;}  
			\Else 
			\State {$flag$ $\gets$ $false$;} 
			\EndIf
			\State {$return$; }
			\EndProcedure	
			\algstore{valid}
		\end{algorithmic}
	\end{algorithm}
	\begin{algorithm}[H]
		\captionsetup{font=scriptsize}
		\caption{\loct Method: Takes $key$ as input and returns the corresponding pair of neighboring $\node$ $\langle n_1, n_2 \rangle$. Initially $n_1$ and $n_2$ are set to $null$.}
		\label{alg:locate}
		\begin{algorithmic}[1]
			\scriptsize
			\algrestore{valid}
			\Procedure{\loct ($key\downarrow, n_1\uparrow, n_2\uparrow$)}{}
			\While{(true)}
			\State {$n_1 \gets read(\head)$;} \label{lin:loc3} 
			\State {$n_2 \gets read(n_1.next)$;} \label{lin:loc4}
			\While{$(read(n_2.val) < key)$} \label{lin:loc5}
			\State {$n_1 \gets n_2$;} \label{lin:loc6}
			\State {$n_2 \gets read(n_2.next)$;} \label{lin:loc7}
			\EndWhile
			\State {$lock.acquire(n_1)$;} \label{lin:loc9}
			\State {$lock.acquire(n_2)$;} \label{lin:loc10}
			\If{($\valid(n_1\downarrow, n_2\downarrow, flag\uparrow$))} \label{lin:loc11}
			\State {return;} \label{lin:loc-ret}
			\Else
			\State {$lock.release(n_1)$;}
			\State {$lock.release(n_2)$;}
			\EndIf
			\EndWhile
			\EndProcedure
			\algstore{locate}
		\end{algorithmic}
	\end{algorithm}
	\begin{algorithm}[H]
		\captionsetup{font=scriptsize}
		\caption{\add Method: $key$ gets added to the set if it is not already part of the set. Returns $true$ on addition.}
		\label{alg:add}
		\begin{algorithmic}[1]
			\scriptsize
			\algrestore{locate}
			\Procedure{\add ($key\downarrow, flag \uparrow$)}{}
			\State {$\loct(key\downarrow, n_1 \uparrow, n_2\uparrow)$;}   \label{lin:add2}
			\If {$(read(n_2.val) \neq key$)} \label{lin:add3}
			\State {$write(n_3, \text{new \node}(key))$;} \label{lin:add4}
			\State {$write(n_3.next, n_2)$;} \label{lin:add5}
			\State {$write(n_1.next, n_3)$;} \label{lin:add6}
			\State {$flag$ $\gets$ $true$;}
			\Else
			\State {$lock.release(n_1)$;}
			\State {$lock.release(n_2)$;}
			\State {$flag$ $\gets$ $true$;}    
			\EndIf
			\EndProcedure
			\algstore{add}
		\end{algorithmic}
	\end{algorithm}
	\begin{algorithm}[H]
		\captionsetup{font=scriptsize}
		\caption{Remove Method: $key$ gets removed from the set if it is already part of the set. Returns $true$ on successful remove otherwise returns $false$.}
		\label{alg:remove}
		\begin{algorithmic}[1]
			\scriptsize
			\algrestore{add}
			\Procedure{\rem ($key\downarrow, flag\uparrow$)}{}
			\State {$\loct(key\downarrow, n_1 \uparrow, n_2\uparrow);$}   \label{lin:rem2}   
			\If {$(read(n_2.val) = key)$} \label{lin:rem3} 
			\State {$write(n_2.marked, true)$;} \label{lin:rem4}
			\State {$write(n_1.next, n_2.next)$;} \label{lin:rem5}
			\State {$flag$ $\gets$ $true$;}\label{lin:rem8}
			\Else
			\State {$flag$ $\gets$ $false$;}
			\EndIf
			\State {$lock.release(n_1)$;} \label{lin:rem6}
			\State {$lock.release(n_2)$;} \label{lin:rem7}
			\State {$return$;}
			\EndProcedure
			\algstore{rem}
		\end{algorithmic}
	\end{algorithm}
	\begin{algorithm}[H]
		\captionsetup{font=scriptsize}
		\caption{\con Method: Returns $true$ if $key$ is part of the set and returns $false$ otherwise.}
		\label{alg:contains}
		\begin{algorithmic}[1]
			\scriptsize
			\algrestore{rem}
			\Procedure{\con ($key\downarrow, flag\uparrow$)}{}
			\State {$n \gets read(\head);$}  \label{lin:con2} 
			\While {$(read(n.val) < key)$} \label{lin:con3} 
			\State {$n \gets read(n.next);$} \label{lin:con4} 
			\EndWhile \label{lin:con5}
			\If {$(read(n.val) \neq key) \vee (read(n.marked))$} \label{lin:cont-tf} \label{lin:con6}
			\State {$flag$ $\gets$ $false$;}
			\Else
			\State {$flag$ $\gets$ $true$;}
			\EndIf
			\State {$return$;}
			\EndProcedure
		\end{algorithmic}
	\end{algorithm}
\end{multicols}

\subsubsection*{\textbf{The Linearization Points of the Lazy List methods}}
\noindent Here, we list the linearization points (\lp{s}) of each method. Note that each method of the list can return either $true$ or $false$. So, we define the $\lp$ for six methods:

\begin{enumerate}
	\item $\add(key, true)$: $write(n_1.next, n_3)$ in Line \ref{lin:add6} of $\add$ method.
	\item $\add(key, false)$: $read(n_2.val)$ in Line \ref{lin:add3} of $\add$ method. 
	\item $\rem(key, true)$: $write(n_2.marked, true)$ in Line \ref{lin:rem4} of $\rem$ method.
	\item $\rem(key, false)$: $read(n_2.val)$ in Line \ref{lin:rem3} of $\rem$ method.
	\item $\con(key, true)$: $read(n.marked)$ in \lineref{cont-tf} of $\con$ method.
	\item $\con(key, false)$: $\lp$ is the last among the following lines executed. There are three cases here: 
	\begin{itemize}
		\item [(a)] $read(n.val) \neq key$ in \lineref{cont-tf} of $\con$ method is the $\lp$, in case of no concurrent $\add(key, true)$.
		\item [(b)] $read(n.marked)$ in \lineref{cont-tf} of $\con$ method is the $\lp$, in case of no concurrent $\add(key, true)$ (like the case of \con(key, true)).
		\item [(c)] in case of concurrent $\add(key, true)$ by another thread, we add a dummy event just before Line \ref{lin:add6} of $add(key, true)$. This dummy event is the $\lp$ of $\con$ method if: (i) if in the post-state of $read(n.val)$ event in \lineref{cont-tf} of \con method, $n.val \neq key$ and $write(n_1.next, n_3)$ (with $n_3.val = key$) in Line \ref{lin:add6} of \add method executes before this $read(n.val)$. (ii) if in the post-state of $read(n.marked)$ event in \lineref{cont-tf} of \con method, $n.marked = true$ and $write(n_1.next, n_3)$ (with $n_3.val = key$) in Line \ref{lin:add6} of \add method executes before this $read(n.marked)$.
		
	\end{itemize}
\end{enumerate}
\subsection{Lock-free List PCode}  
\begin{multicols}{2}
	\tiny
	\ignore{
		\begin{algorithm}[H]
			\captionsetup{font=scriptsize}
			\caption{\valid Method: Takes two vertices, $n_1, n_2$, each of type $List$ as input and validates for presence in vertex list and returns $true$ or $false$.} 
			\label{validate:Vertex-lf}
			\begin{algorithmic}[1]
				\scriptsize
				\Procedure{\valid($n_1\downarrow, n_2\downarrow, flag \uparrow$)}{} 
				\If {$(read(n_1.marked) = false) \wedge (read(n_2.marked) = false) \wedge (read(n_1.next) = n_2) $} \label{lin:valv2}
				\State {$flag \gets true;$}  \label{lin:valv3} \hspace{.5cm} {   // validation successful} 
				\Else 
				\State {$flag \gets false;$}  \label{lin:valv5} \hspace{.5cm} {    // validation fails} 
				\EndIf 
				\State {$return$;} \hspace{.5cm}  {     //return flag}
				\EndProcedure
				\algstore{valv}
			\end{algorithmic}
		\end{algorithm}
	}
	\vspace{-0.6cm}
	\begin{algorithm}[H]
		\captionsetup{font=scriptsize}
		\caption{\lfloct Method: Takes $key$ as input and returns the corresponding pair of neighboring $\node$ $\langle n_1, n_2 \rangle$. Initially $n_1$ and $n_2$ are set to $null$.}
		\label{locate:Vertex-lf}
		\begin{algorithmic}[1]
			\scriptsize
			\Procedure{\lfloct($key\downarrow, n_1\uparrow, n_2\uparrow $)}{}
			\State{bool snip;} 
			\State{List pred, curr, succ;}
			\While{($true$)} \label{lin:search_again}
			\State {$pred \gets read(\head)$;} \label{lin:locv3}
			\State {$curr \gets read(n_1.next)$;} \label{lin:lf-readn1}
			\While{$(true)$}
			\State{succ $\gets$get\_marked\_ref(curr);}
			\While{($is\_marked\_ref(curr))$} \label{lin:locv5} 
			\State{snip $\gets$ CAS(pred.next, curr, succ, false, false);}
			\If{($!snip$)}
			\State{goto \lineref{search_again};}
			\EndIf
			\State{curr $\gets$ succ}
			\State{succ $\gets$ get\_marked\_ref(curr);}
			\EndWhile
			\If{(curr.val $\geq$ key)}
			\State{$n_1 \gets$ pred;}
			\State{$n_2 \gets$ curr;}
			\EndIf
			\State{pred $\gets$ curr}
			\State{curr $\gets$ succ;}\label{lin:lf-readn1-next}
			\EndWhile
			\EndWhile
			\EndProcedure
			\algstore{locv}
		\end{algorithmic}
	\end{algorithm}
	\vspace{-0.6cm}
	
	\begin{algorithm}[H]
		\captionsetup{font=scriptsize}
		\caption{\lfadd  Method: $key$ gets added to the set if it is not already part of the set. Returns $true$ on addition.}
		\label{alg:nadd:Vertex-lf}
		\begin{algorithmic}[1]
			\algrestore{locv}
			\scriptsize
			\Procedure{\lfadd ($key\downarrow, flag\uparrow$)}{}
			\While{($true$)}
			\State {$\lfloct(key\downarrow,n_1\uparrow,n_2\uparrow )$;} 
			\label{lin:naddv2}
			\If {$(read(n_2.val) = key$)}  \label{lin:addv3}  \hspace{.5cm} {// key present} 
			\State{$flag \gets true$;}
			\State {$return$;}
			\Else
			\State {$write(v_3, \text{new $GNode$}(key))$;} \label{lin:addv4} \hspace{.5cm} {// new GNode created} 
			\State {$write(v_3.next, n_2)$;}  \label{lin:addv5} 
			\State {/* Changing the $n_1.next$ to $v_3$ if both $n_2$ and $n_1$ are unmarked*/ }
			\If{($CAS(n_1.next, n_2, v_3, false, false)$)} \label{lin:naddedv}
			\State{$flag \gets true$;}
			\State {$return$;}  \hspace{.5cm} {//added in the vertex list} 
			\EndIf
			\EndIf
			\EndWhile
			\EndProcedure
			\algstore{addv}
		\end{algorithmic}
	\end{algorithm}
	\vspace{-0.6cm}
	\begin{algorithm}[H]
		\captionsetup{font=scriptsize}
		\caption{\lfrem Method: $key$ gets removed from the set if it is already part of the set. Returns $true$ on successful remove otherwise returns $false$.}
		\label{alg:remove:Vertex-lf}
		\begin{algorithmic}[1]
			\algrestore{addv}
			\scriptsize
			\Procedure{\lfrem($key\downarrow,flag\uparrow$)}{}
			\While{($true$)} \label{lin:start-again}
			\State {$\lfloct(key\downarrow,n_1\uparrow,n_2\uparrow )$}    \label{lin:remv2}
			\If {$(read(n_2.val) = key)$} \label{lin:nremv}{ \hspace{.35cm} // present} 
			\State {/* Changing the $n_2.marked$ to $true$ if both $n_2$ and $n_1$ are unmarked*/ }
			\If{($!CAS(n_1.next, n_2, n_2, false, true)$)} \label{lin:lf-logical-marked} {// logically removed}
			\State{go to \lineref{start-again};}
			\EndIf 
			\State {/* Changing the $n_1.next$ to $n_2.next$ if both $n_1$ and $n_2.next$ are unmarked*/ }
			\State{$CAS(n_1.next, n_2, n_2.next, false, false)$;}  {// physically removed}
			\Else 
			\State {$flag \gets false;$}  \label{lin:nremv12}\hspace{.35cm} {// not present}
			\EndIf  
			\State {$return$;}\hspace{.35cm} {//return flag}
			\EndWhile
			\EndProcedure
			\algstore{remv}
		\end{algorithmic}
	\end{algorithm}
	\begin{algorithm}[H]
		\captionsetup{font=scriptsize}
		\caption{\wfcon Method: Returns $true$ if $key$ is part of the set and returns $false$ otherwise.}
		\label{contains:list-wf}
		\begin{algorithmic}[1]
			\algrestore{remv}
			\scriptsize
			\Procedure{\wfcon ($key\downarrow, flag\uparrow$)}{}
			\State {$v \gets read(\head);$}   \label{lin:conv2}
			\While {$(read(v.val) < key)$} \label{lin:conv3}
			\State {$v \gets read(v.next);$} \label{lin:con-lf-lp}
			\EndWhile  \label{lin:conv5}
			\If{ $((read(v.val) = key$) $\wedge$ ($is\_marked\_ref(v) = false))$}\label{lin:conv6-lf} 
			\State {$flag \gets true;$}  \label{lin:conv7}
			\Else
			\State {$flag \gets flase;$}  \label{lin:conv9}
			\EndIf
			\State {$return$;} 
			\EndProcedure
		\end{algorithmic}
	\end{algorithm}
\end{multicols}

\subsection*{Linearization Points: based on \lf List }
\noindent Here, we list the \lp{s} of each method. As each method of the list can return either $true$ or $false$ except \naddv. So, we define the $\lp$ based on that and given bellow:

\begin{enumerate}
	\item $\lfadd(key, true)$:  Successful CAS in Line \ref{lin:naddedv} of $\lfadd$ method.
	\item $\add(key, false)$: $read(n_2.val)$ in Line \ref{lin:addv3} of $\add$ method. 
	\item $\lfrem(key, true)$:  Successful CAS for logical marked in the \lineref{lf-logical-marked} of $\lfrem$ method.
	\item $\lfrem(key, false)$: The LP be in the \lineref{lf-readn1} or \lineref{lf-readn1-next} in the $\lfloct$ method.
	\item $\wfcon(key, true)$: $read(v.next);$ in the \lineref{conv6-lf} of $\wfcon$ method.
	\item $\wfcon(key, false)$: $\lp$ is the last among the following lines executed. There are two cases depending on concurrent $\lfadd$ method: 
	\begin{itemize}
		\item [(a)] with no successful concurrent $\lfadd(key, true)$:  $read(v.next);$ in the \lineref{conv6-lf} of $\wfcon$ method.
		\item [(b)] with successful concurrent $\lfadd(key, true)$: we add a dummy event just before \lineref{naddedv} of \lfadd method. This dummy event is the \emph{\lp} of the \wfcon method.
	\end{itemize}
\end{enumerate}

\end{document}